\documentclass[twocolumn,showpacs,preprintnumbers,amsmath,amssymb]{revtex4}
\usepackage[pdftex]{color,graphicx}

\usepackage{graphicx}
\usepackage{dcolumn}
\usepackage{bm}

\begin{document}

\title{Quantitative vectorial spin analysis in ARPES: Bi/Ag(111) and Pb/Ag(111)}
\author{Fabian Meier$^{1,2}$}
\author{Hugo Dil$^{1,2}$}
\author{Jorge Lobo-Checa$^{3}$}
\author{Luc Patthey$^{2}$}
\author{J\"urg Osterwalder$^{1}$}
\affiliation{$^{1}$Physik-Institut, Universit\"at Z\"urich, Winterthurerstrasse 190, CH-8057 Z\"urich, Switzerland \\ $^{2}$ Swiss Light Source, Paul Scherrer Institut, CH-5232 Villigen, Switzerland \\ $^{3}$Departement Physik, Universit\"at Basel, Klingelbergstrasse 82, CH-4056 Basel, Switzerland}
\date{\today}

\begin{abstract}
The concept of vectorial spin analysis in spin and angle resolved photoemission is illustrated in this report. Two prototypical systems, Bi/Ag(111)$({\sqrt{3}\times\sqrt{3})R30^{\circ}}$ and Pb/Ag(111)$({\sqrt{3}\times\sqrt{3})R30^{\circ}}$, which show a large Rashba type
spin-orbit splitting, were investigated by means of spin and angle resolved photoemission. The spin polarization vectors of individual bands are determined by a two-step fitting routine.
First, the measured intensities are fitted with an appropriate number of suitable peaks to quantify the contributions of the individual bands, then the measured spin polarization curves are fitted by varying for each band the polarization direction and its magnitude. We confirm that the surface states experience a large spin splitting. Moreover, we find that all surface state bands are 100 percent spin polarized, and that for some states spin polarization vectors rotate out of the surface plane.
\end{abstract}


\maketitle

Methods that allow to control and measure the electron spin, or the average of a certain number of spins, have received growing attention in the last few years. In spintronics, the spin field-effect transistor as proposed by Datta and Das \cite{datta}, which relies on the Rashba-Bychkov effect \cite{rashba, bihlmayer1} (henceforth Rashba effect) to manipulate electron spins by an electric field, is one of the key elements. Spin rotation is achieved by a field and momentum dependent spin splitting of bands in a two-dimensional electron gas. While actual devices are currently realized in semiconductor heterostructures \cite{spinfet}, fundamental issues can be more easily studied in two-dimensional metallic systems involving heavy metal atoms, where spin splittings are much larger \cite{hochstrasser,koroteev}. Very recently, a new class of material systems was identified where this effect is even further enhanced, among them the two surface alloys Bi/Ag(111)$({\sqrt{3}\times\sqrt{3})R30^{\circ}}$ and Pb/Ag(111)$({\sqrt{3}\times\sqrt{3})R30^{\circ}}$ \cite{ast,pacile}, referred to as Bi/Ag(111) and Pb/Ag(111) henceforth. Due to an additional reduction of the surface symmetry caused by the $({\sqrt{3}\times\sqrt{3})R30^{\circ}}$ surface reconstruction and due to a slight corrugation of the surface \cite{bihlmayer},
the size of the Rashba type spin-orbit induced spin splitting is about one order of magnitude larger than what is observed for the Au(111) surface state \cite{lashell,moritz1}.\\
We have performed spin and angle resolved photoemission spectroscopy (SARPES) on Bi/Ag(111) and Pb/Ag(111). Furthermore we present a novel two-step fitting routine for the determination of the three-dimensional spin polarization vector of individual bands, thus revealing the complete spin structure of Rashba systems in momentum space.
In order to illustrate our results and the power of SARPES in combination with an adequate model for the data analysis, this paper is arranged as follows: In the first section, the theoretical aspects of the vectorial spin analysis are outlined. In the second section some subsequently relevant physics of the studied systems is introduced, followed by an experimental section. In the last section, the experimental results are presented and discussed.

\section{Spin analysis}

\begin{figure}[b]
\begin{center}
\includegraphics[width=0.48\textwidth]{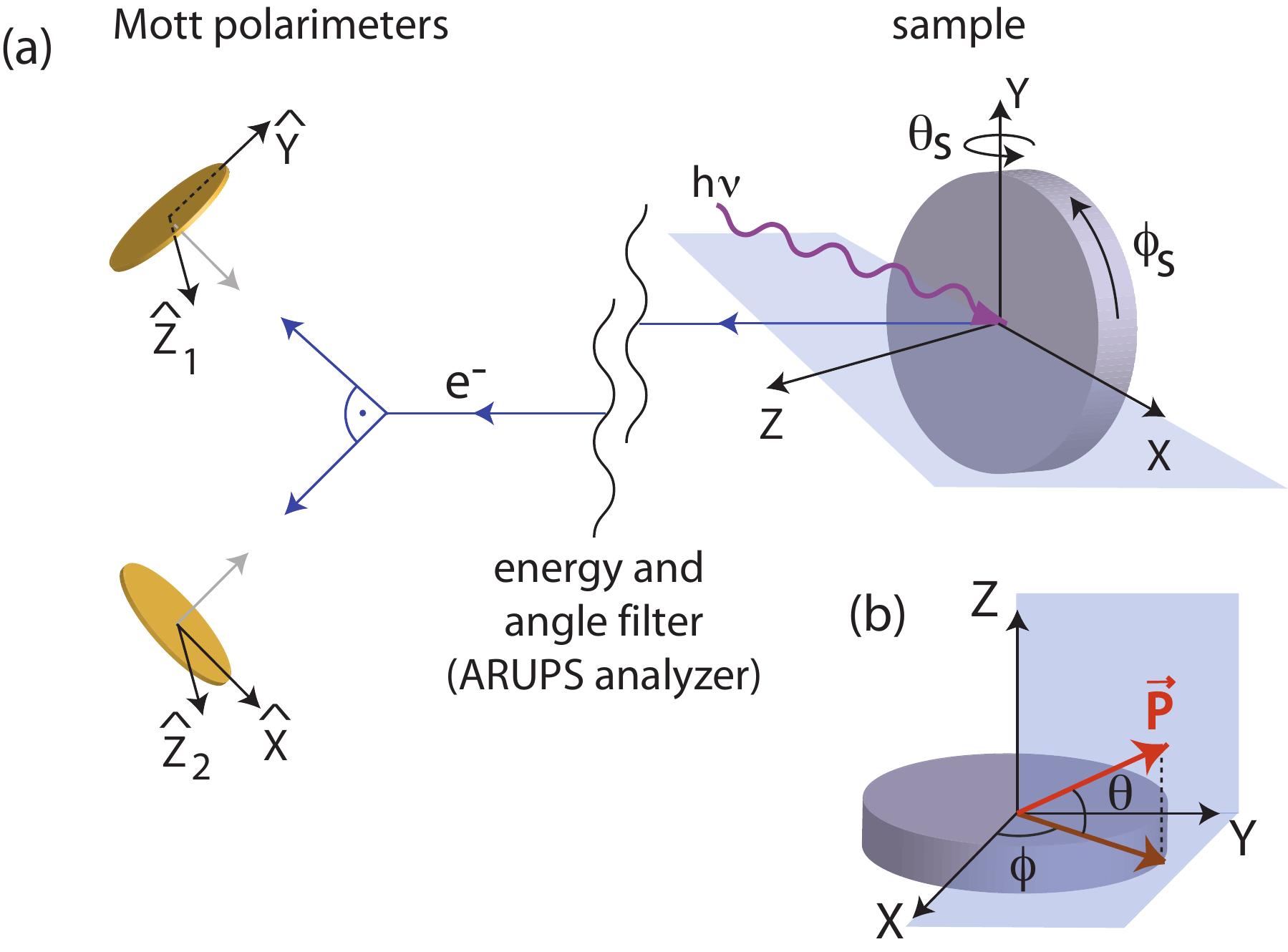}
\caption{{(color online) (a) Schematic illustration of the experimental setup, showing on the right hand side the sample geometry and on the left hand side the three-dimensional Mott polarimeter with two orthogonal gold foils. The coordinate system given by the Mott polarimeters deviates from the sample coordinates through a rotation matrix $T$. (b) Illustration of the spin polarization vector in the sample coordinate system.
}}
\label{Fig1}
\end{center}
\end{figure}

\begin{figure}[htb]
\begin{center}
\includegraphics[width=0.48\textwidth]{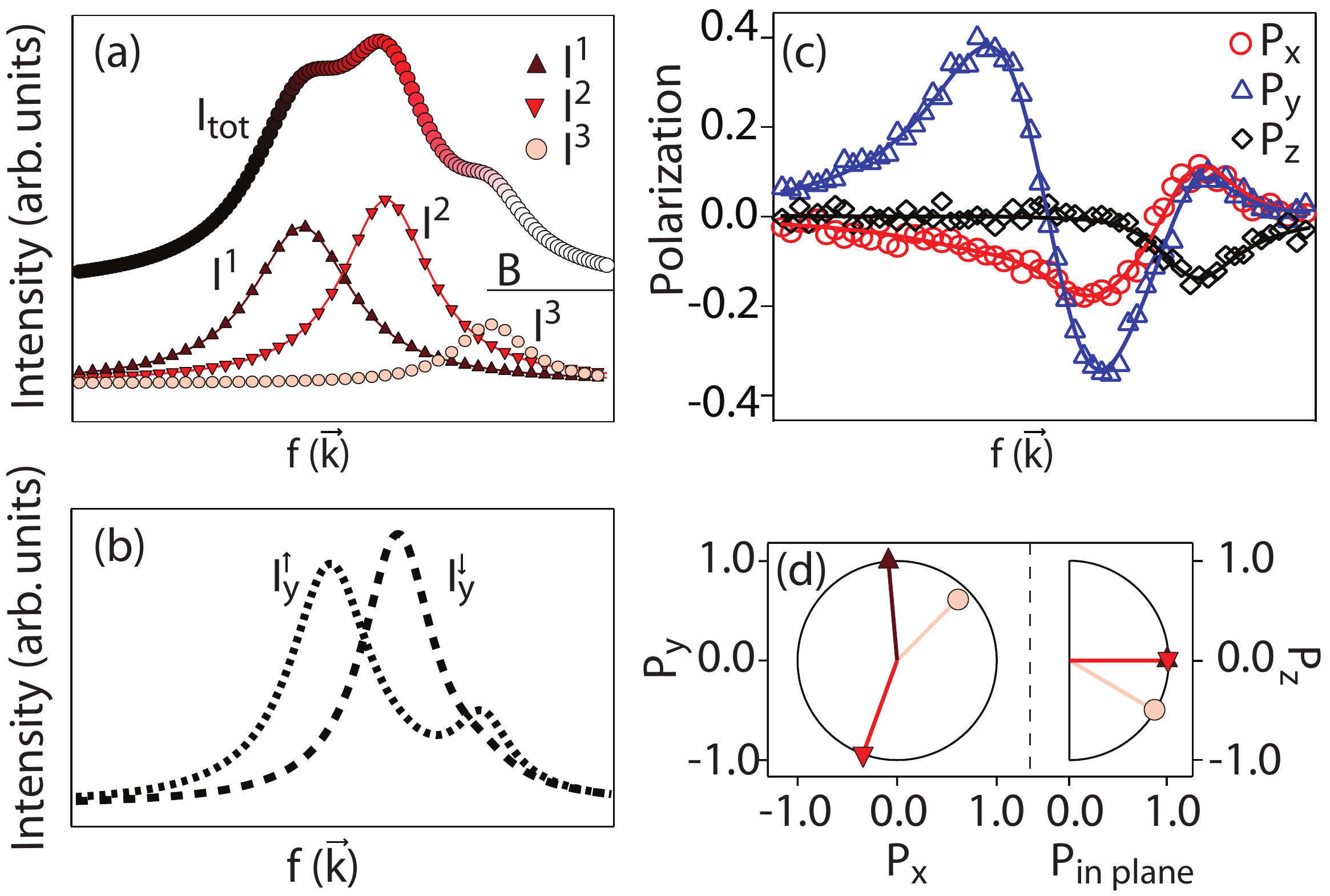}
\caption{{(color online) Illustration of the vectorial spin analysis with artificial data. (a) Spin integrated intensities along one momentum dependent coordinate $f(\vec{k})$ and the peaks extracted from the intensity fit. (b) Spin resolved spectra for the $y$ component, based on an arbitrarily defined spin polarization vector for each band. (c) Spin polarization curves (symbols) for all three spin components, obtained from curves like those given in (b). The lines show the spin polarization curves obtained by using the two-step fitting routine. (d) In plane (left) and out of plane (right) components of the spin polarization vectors of the different bands as obtained from the spin polarization fit. The symbols correspond to those in (a).}}
\label{Fig2}
\end{center}
\end{figure}

In spin and angle resolved photoemission, a typical data set measured with a three-dimensional Mott polarimeter consists of intensity data $I_{\hat{\alpha}}^{1,2}(\vec{k},E)$, with \linebreak $\hat{\alpha}=\hat{x},\hat{y},\hat{z}$ the coordinates of the three orthogonal directions fixed by the Mott scattering planes. Fig. \ref{Fig1}(a) shows the experimental setup and illustrates the difference between the sample coordinates and the coordinates fixed by the Mott detectors. Each direction $\hat{\alpha}$ represents the normal to a scattering plane, defined by the electron incidence direction on the gold foil and two detectors for backscattered electrons. The right-left asymmetry of intensities $I_{\hat{\alpha}}^{1}$ and $I_{\hat{\alpha}}^{2}$ on these two detectors is proportional to the spin polarization component $P_{M\hat{\alpha}}$ of the incident electron beam \cite{kessler,sens}: $P_{M\hat{\alpha}}\nolinebreak=\nolinebreak(1/S)\nolinebreak\cdot\nolinebreak(\nolinebreak I_{\hat{\alpha}}^{1}\nolinebreak-\nolinebreak I_{\hat{\alpha}}^{2}\nolinebreak)\nolinebreak/\nolinebreak I_{\hat{\alpha}}$ with $I_{\hat{\alpha}}=I_{\hat{\alpha}}^{1}+I_{\hat{\alpha}}^{2}$ and $S$ the Sherman function \cite{sherman}, which accounts for the spin detection efficiency.
The spin polarization $\vec{P}$ in the sample coordinates $x,y,z$ can be calculated using a rotation matrix $T$, which depends on the manipulator angles $\theta_{S}$ and $\phi_{S}$ \cite{moritz}: $\vec{P}=T\vec{P}_{M}$. \\
We now have the measured intensities $I_{\hat{\alpha}}^{1,2}$ as well as the spin polarization data $\vec{P}_{M}$ and $\vec{P}$, all as a function of $\vec{k}$ and $E$, but do not possess any direct information about the spin polarization vector of the individual bands contributing to the measured spin polarization. 
Typically, the data analysis of SARPES studies ends at this point. Here we present an approach to gain deeper insight into the vectorial spin polarization of the system under investigation.
It is a two step fitting routine that determines the magnitude and the direction of the spin polarization vector of each individual band.

\subsection*{Two-step fitting routine}
The measured polarization data are compromised by the overlap of adjacent peaks as well as by the background and do not directly reveal the spin polarization vector of each individual band. In order to overcome this limitation a quantitative vectorial spin analysis that allows for the determination of the spin polarization vector of each band is needed.
\newline
In the analysis, we first fit the measured intensity data $I_{tot}=\sum_{\hat{\alpha}}I_{\hat{\alpha}}$ with an appropriate number of suitable peaks $I^{i}$ (e.g. Lorentzians), corresponding to
the individual bands, plus a background $B$, which we assume to be constant and non polarized throughout this work,
\begin{eqnarray}
I_{tot}=\sum_{i=1}^{n} I^{i} + B,
\label{Ifit}
\end{eqnarray}
with $n$ the number of bands. This is illustrated in Fig. \nolinebreak \ref{Fig2} \nolinebreak (a) with an artificially generated data set.
In the second step of the fitting routine, we start by assigning a spin polarization vector $\vec{P}^{i}$ to each band, defined as
\begin{eqnarray}
\vec{P}^{i}=(P_{x}^{i},P_{y}^{i},P_{z}^{i})=c_{i}(\cos\theta_{i}\cos\phi_{i},\cos\theta_{i}\sin\phi_{i},\sin\theta_{i})
\label{PolVecForm}
\end{eqnarray}
with $0\leq c_{i} \leq 1$.
The scalar $c_{i}$ defines the magnitude of the spin polarization vector of band $i$. The angles $\theta_{i}$ and $\phi_{i}$ are defined according to Fig. \nolinebreak \ref{Fig1} \nolinebreak (b). Now one can determine the spin resolved spectra for each peak $i$ by using
\begin{eqnarray}
I_{\alpha}^{i;\uparrow,\downarrow}=I^{i}(1\pm P_{\alpha}^{i})/6, \ \ \alpha=x,y,z,
\end{eqnarray}
with the $+$ and the $-$ corresponding to $\uparrow$ and $\downarrow$, respectively, meaning spin parallel or antiparallel to the direction $\alpha$. From the spin resolved spectra of the different bands the entire spin resolved spectra can be calculated,
\begin{eqnarray}
I_{\alpha}^{\uparrow,\downarrow}=\sum_{i=1}^{n}I_{\alpha}^{i;\uparrow,\downarrow}+B/6,
\end{eqnarray}
where the background is divided equally between the different spatial directions. Fig. \nolinebreak \ref{Fig2} \nolinebreak (b) gives an example of the spin resolved intensity spectra $I_{y}^{\uparrow,\downarrow}$ for the $y$ direction.
The spin polarization of each spatial component $\alpha$ can be obtained from
\begin{eqnarray}
P_{\alpha}=\frac{I_{\alpha}^{\uparrow}-I_{\alpha}^{\downarrow}}{I_{\alpha}^{\uparrow}+I_{\alpha}^{\downarrow}}.
\end{eqnarray}
This yields the spin polarization spectra for a certain set of parameters $\theta_{i}$, $\phi_{i}$ and $c_{i}$. The second step is concluded by fitting the measured polarization data by varying the angles and the magnitudes of the spin polarization vectors for all peaks $i$. The spin polarization spectra resulting from this second step are illustrated in Fig. \ref{Fig2} (c)  (solid lines) for our artificially generated data set. The obtained parameters reveal the spin polarization vectors of the individual bands in the measured band structure. They are displayed in Fig. \ref{Fig2} \nolinebreak (d), where the left hand panel shows the in-plane spin polarization components and the right hand panel the out-of-plane spin polarization component. In the following, the display type of Fig. \ref{Fig2} \nolinebreak (d) will be used to illustrate the spin polarization vectors, but the scaling of the unit circle will be left out for simplicity.
\newline
Thus, in principle, by using the two-step fitting routine described above, the full information about the spin structure of a given system can be obtained.

\section{Rashba effect}

In band theory, it is usually taken for granted that the space inversion symmetry and the time reversal symmetry
are fulfilled. This results in a spin degeneracy of the well known Bloch states \cite{kittel},
\begin{eqnarray}
E(\vec{k}, \uparrow)=E(\vec{k}, \downarrow).
\end{eqnarray} 
This degeneracy can be lifted if either the space inversion symmetry or the time inversion symmetry is broken.
The former is the case in crystals which lack an inversion symmetry centre in the unit cell (Dresselhaus effect) \cite{dresselhaus,parmenter} and at interfaces or surfaces, where it is referred to as the Rashba effect  \cite{rashba}.
The Rashba effect is a result of the spin orbit coupling and can be described by
the following Hamiltonian \cite{winkler},
\begin{eqnarray}
H_{SOI}=-\frac{\hbar^{2}}{4m^{2}c^{2}}(\vec{\nabla}V\times \hat{\vec{p}}\ )\cdot\vec{\sigma},
\label{Hsoi}
\end{eqnarray}
where $V$ is the electronic potential, $\hat{\vec{p}}$ the momentum operator and $\vec{\sigma}$ the vector of Pauli matrices.
As a consequence, the bands are split and completely spin polarized. In the case of a two dimensional free electron gas the dispersion relation is given by
\begin{eqnarray}
E^{\pm}(\vec{k})=E_{0}+\frac{\hbar^{2}|\vec{k}|^{2}}{2m^{*}}\pm \alpha_{R}|\vec{k}|,
\end{eqnarray} 
where $\alpha_{R}$ is the Rashba parameter, which determines the size of the splitting. 
It depends on the atomic spin-orbit interaction and on the surface potential gradient, i.e. $\alpha_{R}=\alpha_{A}\alpha_{V}$ \cite{petersen}. The Rashba parameter can be related to the wave number offset $k_{0}$ of the band extremum from the surface Brillouin zone (SBZ) center,
\begin{eqnarray}
\alpha_{R}=\frac{\hbar^{2}k_{0}}{m^{*}}.
\label{alpha}
\end{eqnarray} 
The Rashba energy, which is given by
\begin{eqnarray}
E_{R}=\frac{\hbar^{2}k_{0}^{2}}{2m^{*}},
\label{Erashba}
\end{eqnarray}
describes the energy difference between the extrema and the crossing of the spin split bands.
\newline
It should be noted that although the states are spin polarized the surface remains non magnetic due to the time inversion symmetry.

\subsection*{Bi/Ag(111) and Pb/Ag(111)}
The surface states of the two surface alloys Bi/Ag(111) and Pb/Ag(111) show a strongly enhanced Rashba effect \cite{ast,ast2}. It was qualitatively shown in a nearly free electron (NFE) model calculation \cite{premper}, that this enhancement is due to a further reduction of the symmetry compared to the sole presence of the surface, caused by an additional in plane inversion asymmetry. In the surface alloys Bi/Ag(111) and Pb/Ag(111), the in plane symmetry is altered due to the $(\sqrt{3}\times\sqrt{3})R30^{o}$ surface reconstruction and the in-plane potential is modulated, as there are comparably light atoms (Ag) surrounding heavier atoms (Bi or Pb).
As a consequence of the large in-plane potential gradients, the spin polarization vector can be rotated out of the surface plane, where the amount of the rotation depends on the crystallographic direction, the in-plane momentum and the band symmetry \cite{ast,bihlmayer,premper}.
Because of the threefold rotational symmetry, possible out-of-plane components (as well as their in-plane counterparts) will have a $2\pi/3$ periodicity. As pointed out by Bihlmayer et al. \cite{bihlmayer}, the strength of the Rashba effect is further influenced by the corrugation of the surface, and it was shown for the case of Pb/Ag(111), that an artificial reduction of the corrugation reduces the spin splitting. 

\section{Experimental}
The sample preparation was carried out in situ under ultra high vacuum (UHV) conditions with a base pressure better than $2\cdot10^{-10}$ mbar. The Ag(111) crystal was cleaned by multiple cycles of Ar$^{+}$ sputtering and annealing, where the cleanliness was confirmed by the observation of the L-gap surface state. A third of a monolayer of either Pb or Bi was deposited by evaporating the materials from a home made Knudsen effusion cell, with the sample held at 80$^{\circ}$ C. The sample quality was affirmed by low energy electron diffraction (LEED) and ARPES.
\newline
The experiments were performed at room temperature at the Surface and Interface Spectroscopy (SIS) beamline at the Swiss Light Source (SLS) of the Paul Scherrer Institute (PSI) using the COPHEE (the COmplete PHotoEmission Experiment) spectrometer, which is described in much detail elsewhere \cite{moritz}. The data were obtained using horizontally polarized light with a photon energy of 21.2 \nolinebreak eV, 23 \nolinebreak  eV or 24 \nolinebreak  eV, depending on the measurement.
\newline
The COPHEE spectrometer is equipped with two orthogonal Mott polarimeters and can measure the three spatial components of the spin polarization vector for an arbitrary point in reciprocal space. The efficiency of Mott polarimeters is about three orders of magnitude lower than the efficiency of common spin-integrated detectors. The acquired data typically contain around half a million counts on the intensity maxima for each scattering plane for which an accumulation time of around 5 minutes per data point is needed.
The Sherman function was determined to be $S=0.085$.

\section{Results and Discussion}
In this work we experimentally confirm that the Bi/Ag(111) and Pb/Ag(111) surface alloys exhibit a strongly enhanced Rashba type spin-orbit splitting and that the spin polarization vectors can be extracted well by applying the spin analysis routines described above. We deliver evidence that the surface state bands are completely spin polarized and we show a direct experimental observation of the out-of-plane spin polarization component. The experimental results confirm to a certain extent previous theoretical work \cite{bihlmayer}.  

\begin{figure}[htb]
\begin{center}
\includegraphics[width=0.48\textwidth]{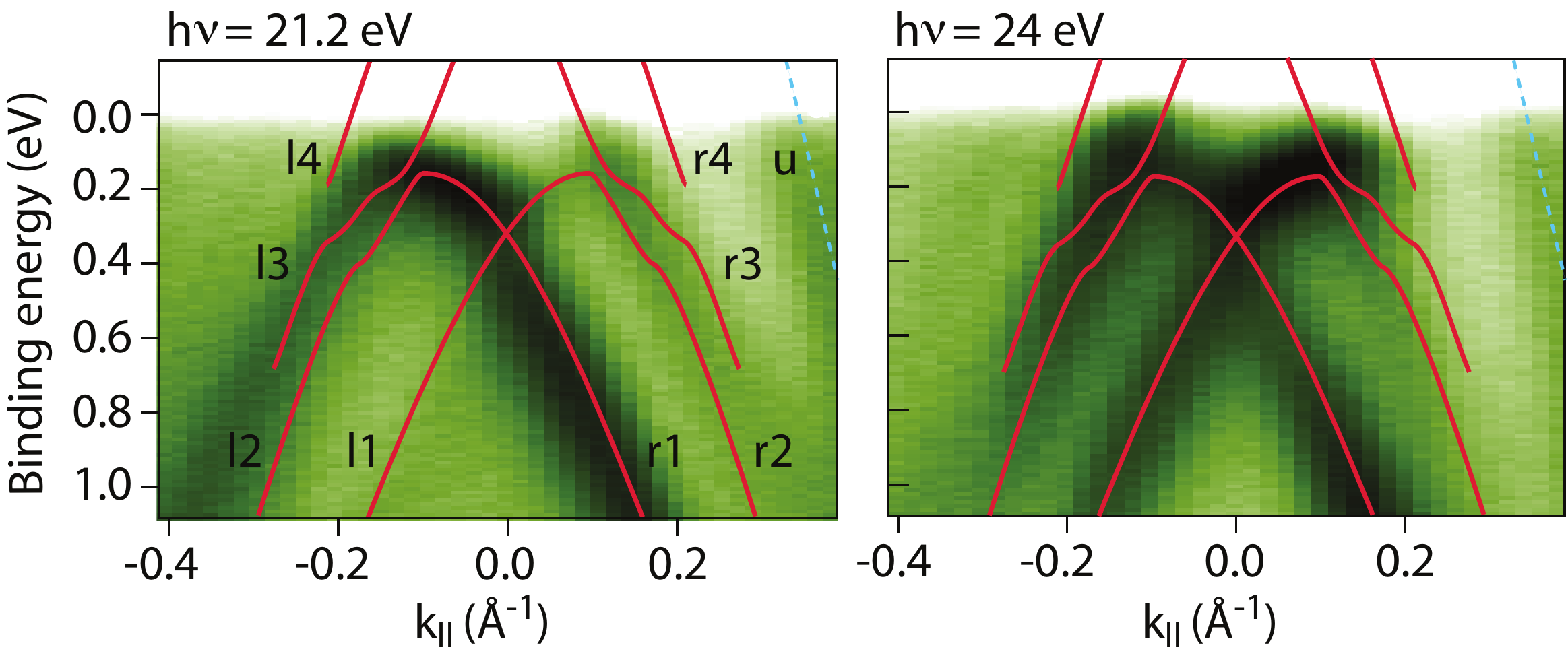}
\caption{{(color online) Cuts through the SBZ along $\bar{\Gamma} \bar{K}$ for the Bi/Ag(111) surface alloy at two different photon energies, $hv$ = 21.2 eV and 24 eV. The calculated band structure (solid lines) is adapted from \cite{bihlmayer}. The labels refer to the different band assignments (see text).}}
\label{Fig3}
\end{center}
\end{figure}

\subsection*{Bi/Ag(111)}

An overview of the band structure and an illustration of the labeling convention of the Bi/Ag(111) surface alloy is given in Fig. \ref{Fig3}. In this figure two spin integrated cuts through the SBZ in the $\bar{\Gamma} \bar{K}$ direction (the crystallographic axes refer to the $\sqrt{3}\times\sqrt{3}R30^{\circ}$ reconstruction throughout this work) measured at $hv$ \nolinebreak = \nolinebreak 21.2 \nolinebreak eV and 24 \nolinebreak eV are shown. The solid lines refer to the band structure from first principles calculations adapted from Ref. \cite{bihlmayer}. In order to get a good correspondence with the experimental data, the energy scale of the adapted band structure is shifted by approximately 180 \nolinebreak meV towards larger binding energies and the $k$ scale is adjusted to fit the experimental results. It shows four surface state bands, which are labeled $l4-r4$. The labeling of the bands is based on their distance from the SBZ center at higher binding energies. This means that the labels $l2$ and $r1$, for example, belong to the same band but at opposite sides of the SBZ center.
The bands $l2/r1$ and $l1/r2$ are Rashba-type spin-split bands derived from a surface state with mostly $sp_{z}$ symmetry. The bands $l4/r3$ and $l3/r4$ are primarily $p_{x,y}$ derived states and can be classified as $m_{j}=1/2$ \cite{bihlmayer}.
The rightmost band (labeled $u$) visible in both cuts through the SBZ in Fig. \ref{Fig3} is a surface umklapp band of the Ag $sp$ band due to the surface reconstruction \cite{pacile} and will not be discussed further.
In the experiment the intensity distribution varies for the different photon energies due to strong final state effects, i.e. at $hv$ \nolinebreak = \nolinebreak 21.2 \nolinebreak eV, $l3$ is more intense than $l2$, while the opposite is the case at $hv$ \nolinebreak = \nolinebreak 24 \nolinebreak eV.

\begin{figure}[b]
\begin{center}
\includegraphics[width=0.41\textwidth]{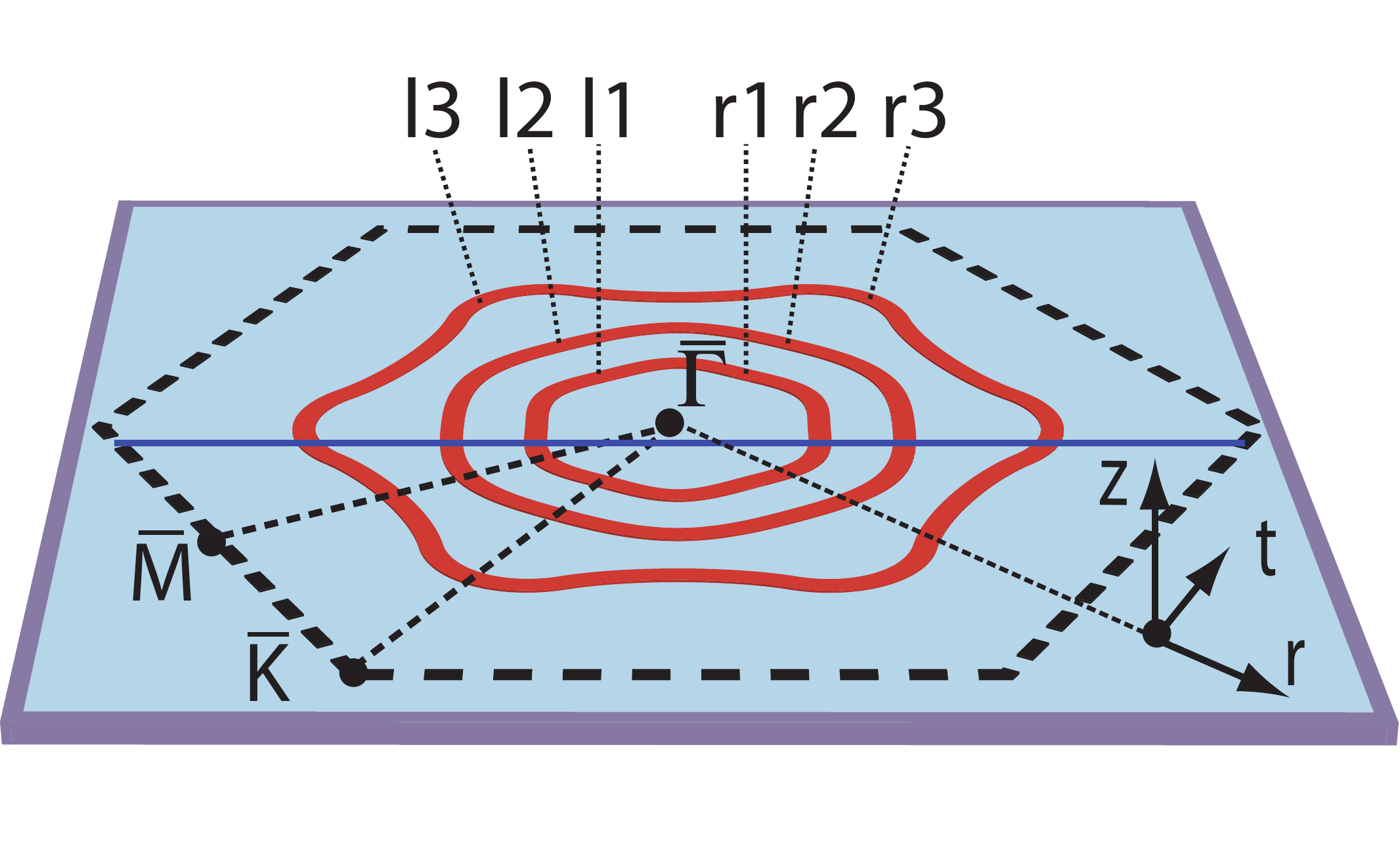}
\caption{{(color online) Schematic constant energy surface map of the Bi/Ag(111) surface alloy for a binding energy of around 0.9 eV. The surface Brillouin zone is given by the thick dashed lines. Constant energy contours are labeled according to the convention described in the text. Due to a small sample tilt, the measured momentum distribution curves (solid line, see following figures) do not pass exactly through the zone center $\bar{\Gamma}$. For the spin analysis, we define for an arbitrary point in reciprocal space the $y$ axis as the tangential component $t$ and the $x$ axis as the radial component $r$; $z$ is the out-of-plane component.}}
\label{Fig4}
\end{center}
\end{figure}

A schematic constant energy surface map for a binding energy below the crossing point of the inner two bands is shown in Fig. \nolinebreak \ref{Fig4} \nolinebreak. The constant energy contours (solid lines) deviate from a circular shape due to the interaction with the crystal lattice \cite{ast}. Due to a small tilt ($<2^{\circ}$) of the sample normal and consequently of the azimuthal sample rotation axis ($\phi_s$, see Fig. \ref{Fig1} (a)), the measured momentum distribution curves (MDCs) do not exactly cut through the $\bar{\Gamma}$-point, as visualized by the horizontal line in Fig. \ref{Fig4}. For the vectorial spin analysis we define, for an arbitrary point in reciprocal space, the radial component $r$ and the tangential component $t$ as the $x$ and $y$ axes of the sample coordinate system, respectively.

\begin{figure*}[htb]
\begin{center}
\includegraphics[width=0.75\textwidth]{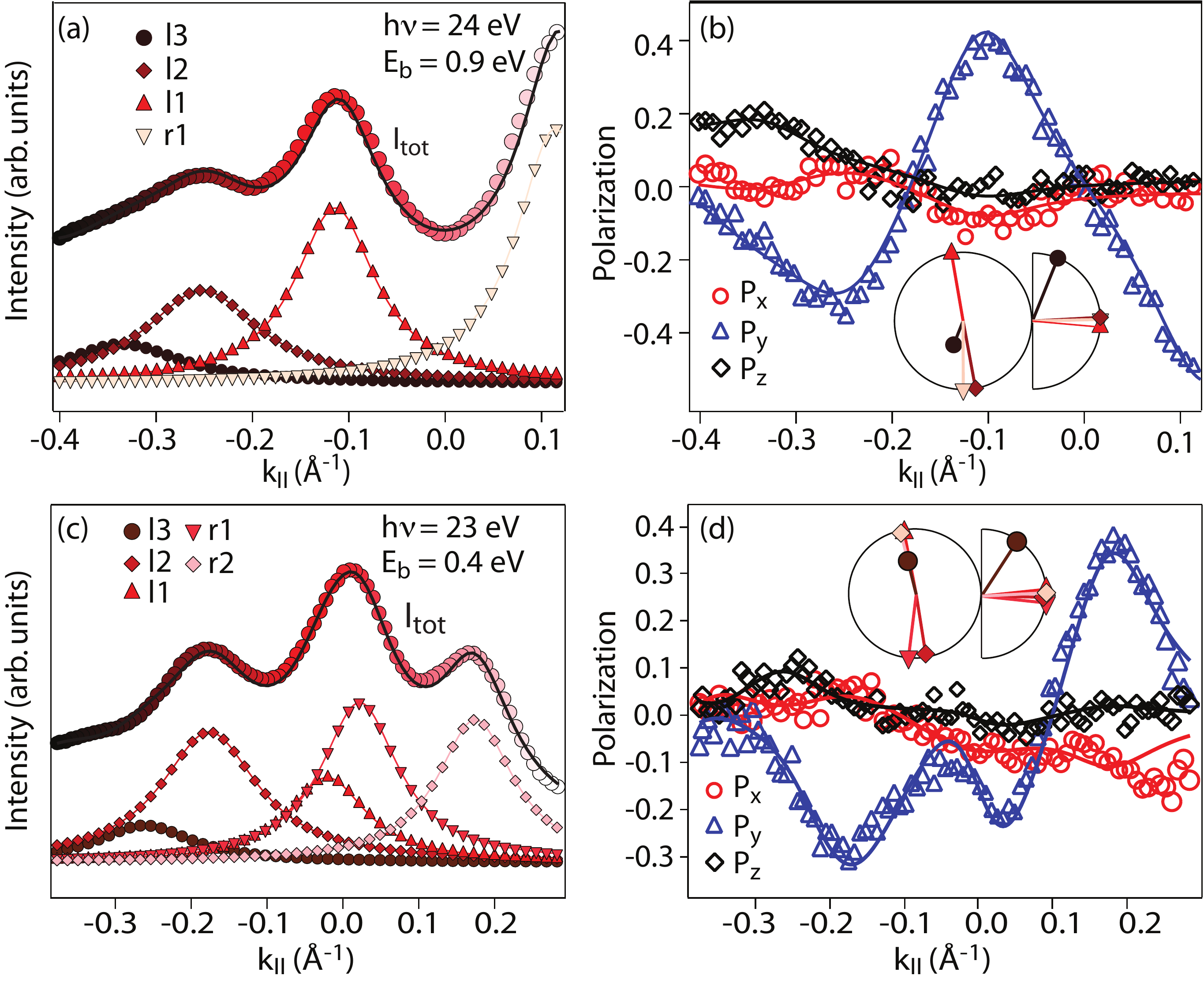}
\caption{{(color online) Momentum distribution curves measured at $E_{b}=0.9$ eV with $hv=24$ eV (top panels) and at $E_{b}=0.4$ eV with $hv=23$ eV (bottom panels) along the $\bar{\Gamma} \bar{K}$ direction. (a) and (c) show the spin integrated intensities and the Lorentzian peaks of the fit. The solid line is the total intensity fit. (b) and (d) show the measured (symbols) and fitted (solid lines) spin polarization curves from the MDC. The statistical errors are smaller than the symbol size. The insets of (b) and (d) visualize the in-plane and out-of-plane spin polarization components obtained from the polarization fit, where the symbols refer to those in (a) and (c), respectively.}}
\label{Fig5}
\end{center}
\end{figure*}

Figs. \ref{Fig5} (a) and (b) show a spin resolved MDC measured in the $\bar{\Gamma} \bar{K}$ direction with $hv=24$ eV at a binding energy $E_{b}=0.9$ eV. The spin polarization vectors of the different bands are determined with the two-step fitting routine described above, and the results are summarized in Table \ref{Table1}.
It is found that the fit parameters $c_{i}$ of the surface states can all be set equal to one, without deteriorating the quality of the fit. This implies that these states are 100 percent spin polarized. 
The fact that the measured spin polarization is smaller than 100 percent finds its origin in the non-polarized background and the overlap of adjacent peaks with different spin polarization.
\newline
The bands $l2/r1$ and $l1/r2$ with mainly $sp_{z}$ character have a spin polarization vector which lies primarily in the surface plane ($\theta < 7^\circ$), with the spin polarization vector pointing approximately in the $\pm y$ direction ($\phi \approx \pm 90^\circ$, roughly tangential to the constant energy contours) as expected from theory. Small deviations $\Delta\phi$ from pure $y$ spin polarization could be observed ($\Delta\phi < 15^{\circ}$) and are assumed to be a ramification of the non-circular constant energy contours in combination with the experimental tilt.
Band $l3$, which carries $p_{x,y}$ character, feels a stronger influence of the in plane potential gradients, and as a consequence the spin polarization vector is rotated out of the surface plane by a significant amount. The rotation angle $\theta$ was determined to be 68$^{\circ}$, which is much larger than the one found for the inner two bands, where a maximum value for $P_{z}$ of about 7$^{\circ}$ was observed both in our experiments and theoretically \cite{ast}. 
\newline
The Rashba parameter $\alpha_{R}$ and the Rashba energy $E_{R}$ of the Bi/Ag(111) structure were extracted using the momentum splitting $\Delta k=0.14$ \AA$^{-1}$\ of the inner two bands at $E_{b}\nolinebreak=\nolinebreak0.9$ \nolinebreak eV and the effective mass $m^{*}\nolinebreak=\nolinebreak-0.35\ m_{e}$ from Ref. \cite{ast}. From Eq. \nolinebreak \ref{alpha} it then follows that $\alpha_{R}\nolinebreak=\nolinebreak3.28$\nolinebreak\ eV\AA\,  and from Eq. \nolinebreak \ref{Erashba} one obtains $E_{R}\nolinebreak=\nolinebreak231$\nolinebreak\ meV. However,  Eq. \nolinebreak \ref{alpha} and Eq. \nolinebreak \ref{Erashba} only hold for a two dimensional free electron gas. In the present case, the situation is more complex and different bands may experience different splitting strengths. Furthermore the dispersion is not a perfect parabola and thus the momentum splitting may vary with binding energy.
\newline
Fig. \ref{Fig5} (c) shows a MDC obtained with $hv=23$ eV at a binding energy of 0.4 eV, i.e. near the crossing point of the inner two bands. This means that the measurement passes through a region with a significant overlap of two bands with opposite spin polarization, and the vectorial spin analysis is necessary to resolve the individual polarization of those bands. The corresponding spin polarization curves are shown in Fig. \ref{Fig5} (d). Again the bands $l2$, $r1$ and $l1$, $r2$ are found to be polarized entirely in-plane, with the spin polarization vector approximately pointing towards the $\pm y$ direction, while the spin polarization of $l3$ is found to be rotated out of the surface plane. The amount of rotation is smaller compared to the measurement performed at 0.9 eV, i.e. 57$^{\circ}$ vs 68$^{\circ}$. 
This finding is consistent with theoretical considerations, where it was found that the amount of out-of-plane spin polarization increases towards the zone boundaries due to the growing influence of the lattice potential and thus of the in-plane potential gradients \cite{premper}.

\begin{figure}[htb]
\begin{center}
\includegraphics[width=0.375\textwidth]{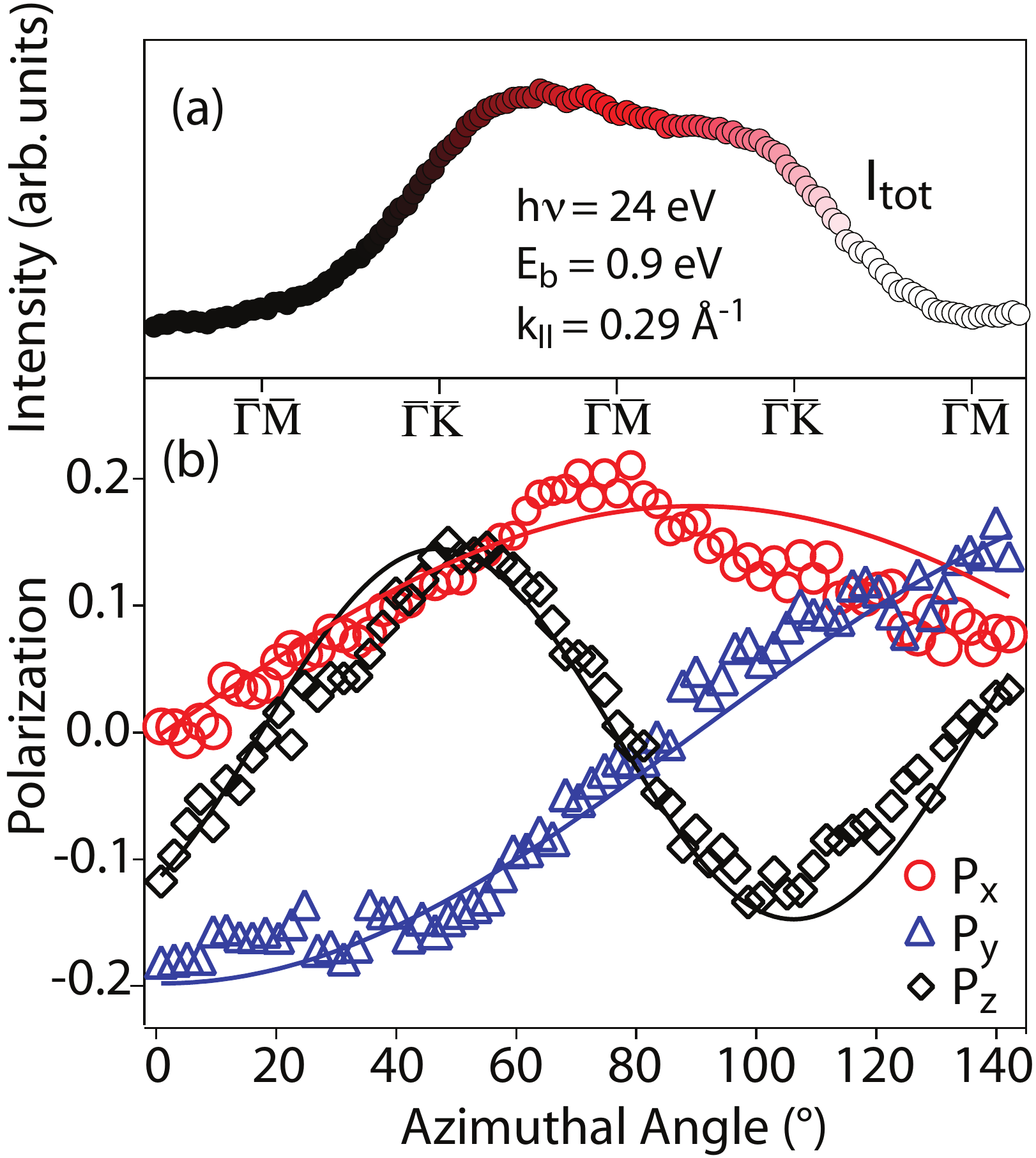}
\caption{{(color online) Azimuthal scan of intensity (a) and spin polarization (b) at $E_{b}=0.9$ eV and $k_{||}=0.29$ \AA$^{-1}$ ($hv=24$ eV). The sine curves (solid lines) are not fits but solely a guide to the eye. The $z$ component of the spin polarization vector shows approximately a sine like behavior with $2\pi/3$ periodicity and extrema in the $\bar{\Gamma}\bar{K}$ direction. The statistical errors are smaller than the symbol size.}}
\label{Fig6}
\end{center}
\end{figure}

As shown above, the strong in-plane potential gradients rotate the spin polarization vector of band $l3$ out of the surface plane by a significant amount. Fig. \ref{Fig6} shows a spin resolved azimuthal scan, where the binding energy and the in plane momentum are fixed at 0.9 eV and 0.29 \AA, respectively, and the crystal is rotated around the surface normal.
We find that the out-of-plane spin polarization component $P_{z}$ of band $l3$ shows a sine like behavior with a $2\pi/3$ periodicity. The $z$ polarization is maximal in the $\bar{\Gamma} \bar{K}$ direction and vanishes in the $\bar{\Gamma} \bar{M}$ direction. As will be shown for the Pb/Ag(111) surface alloy, the spin polarization vector rotates into the surface plane while its total magnitude is not reduced towards $\bar{\Gamma} \bar{M}$. $P_{x}$ and $P_{y}$ refer to a coordinate system that is fixed on the sample for the initial azimuth of the measurement and show a sine and cosine like behavior with $2\pi$ periodicity, in agreement with the assumption that the in-plane part of the polarization vector is tangential to the constant energy contour.

\begin{table}[htb]
\begin{center}
\begin{tabular}{c  c  c  c  c  c }
\hline
\multicolumn{6}{c}{Bi/Ag(111)} \\
\hline
\multicolumn{6}{c}{$E_{b}=0.4$ eV, $\bar{\Gamma} \bar{K}$} \\
$k_{||}$ (\AA$^{-1}$) & -0.26 & -0.18 & -0.02 & 0.02 & 0.18 \\
$\phi$ $(^\circ)$ & 104 $\pm10$ & -81 $\pm2$ & 101 $\pm4$ & -97 $\pm2$ & 104 $\pm1$ \\
$\theta$ $(^\circ)$ & 57 $\pm7$ & -1 $\pm2$ & 6 $\pm5$ & -6 $\pm2$ & 3  $\pm1$\\
\hline
 & $l3$ & $l2$ & $l1$ & $r1$ & $r2$ \\
 \hline
\multicolumn{6}{c}{$E_{b}=0.9$ eV, $\bar{\Gamma} \bar{K}$} \\
$k_{||}$ (\AA$^{-1}$) & -0.34 & -0.25 & -0.11 & 0.12 &  \\
$\phi$ $(^\circ)$  & -111 $\pm17$ & -80 $\pm2$ & 99 $\pm1$ & -90 $\pm1$ &  \\
$\theta$ $(^\circ)$  & 68 $\pm5$ & 3 $\pm2$ & -5 $\pm1$ & 1 $\pm1$ & \\
\hline
\end{tabular}
\end{center}
\caption{Band positions and directions (given by $\theta$ and $\phi$) of the spin polarization vectors of the surface state bands $l3-r2$ for the Bi/Ag(111) surface alloy at $E_{b}=0.4$ eV and $E_{b}=0.9$ eV. The angular errors are estimates resulting from the fitting procedure to the spin polarization spectra.}
\label{Table1}
\end{table}

\subsection*{Pb/Ag(111)}

\begin{figure}[b]
\begin{center}
\includegraphics[width=0.48\textwidth]{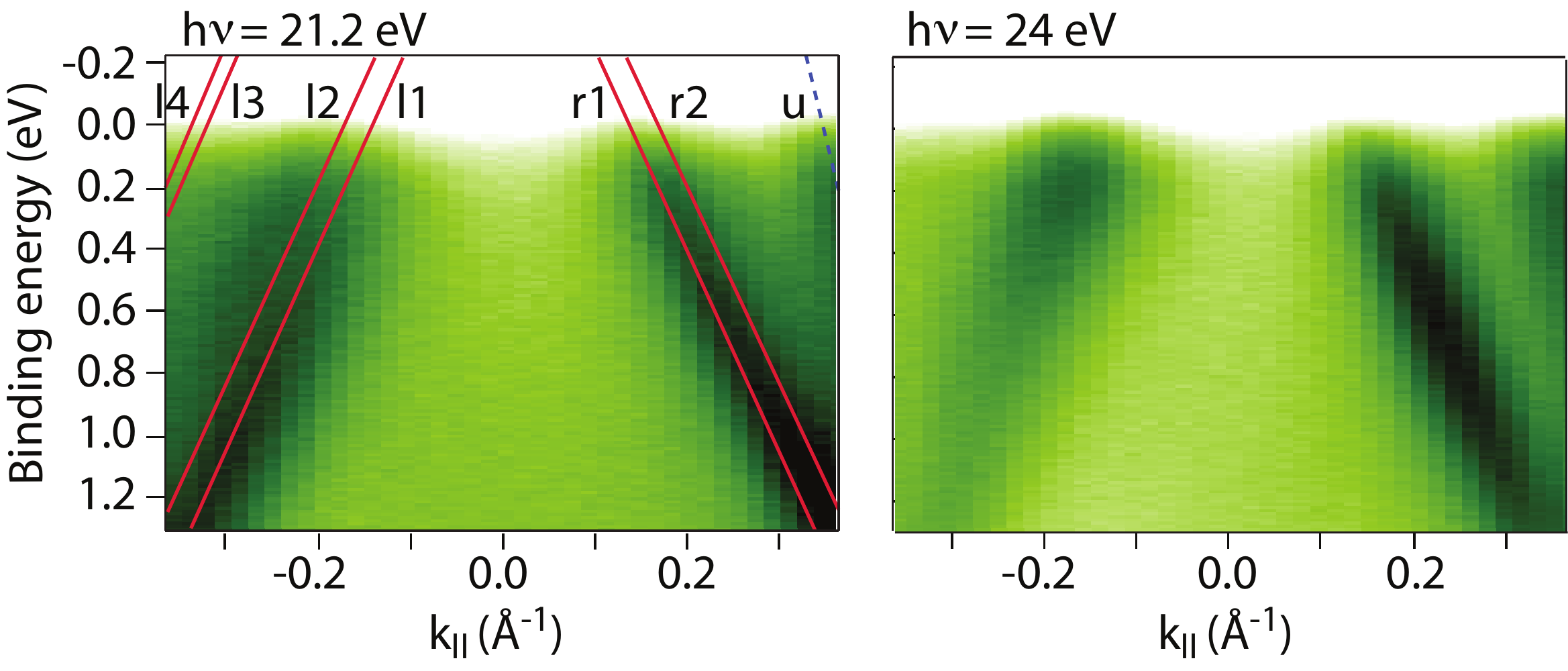}
\caption{{(color online) Spin integrated cuts through the SBZ in the $\bar{\Gamma} \bar{K}$ direction at $h\nu=21.2$ eV and $h\nu=24$ eV. Final state effects are found to be strong, as can readily be seen from the changes in intensity between positive and negative parallel momenta or by comparing the measurements at different photon energies. The solid lines represent the different surface state bands at their approximate positions (according to Ref. \cite{pacile}). The dotted line corresponds to the surface umklapp band of the Ag $sp$ band. The labels are used to mark the different bands.}}
\label{Fig7}
\end{center}
\end{figure}

\begin{figure*}[htb]
\begin{center}
\includegraphics[width=0.75\textwidth]{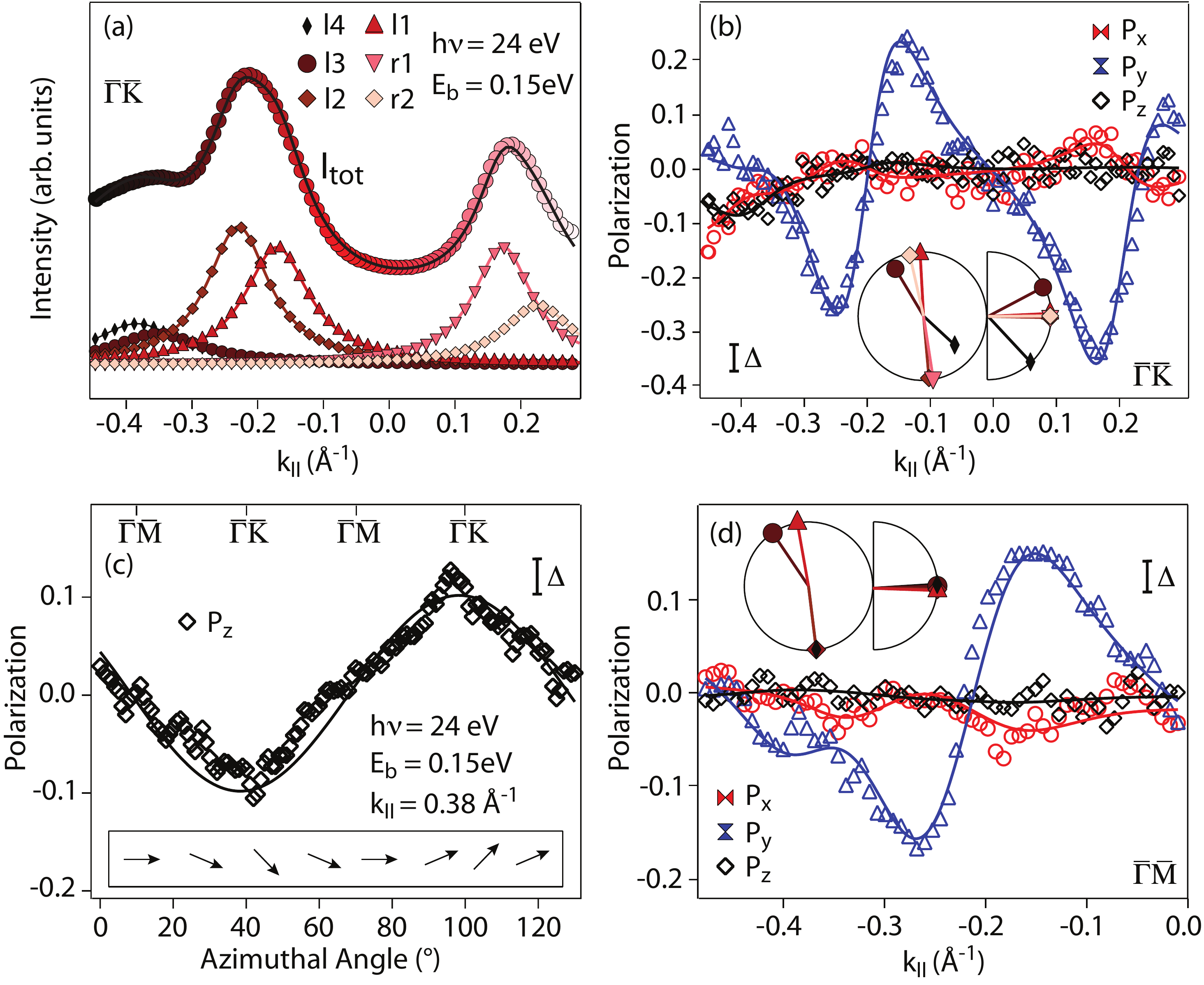}
\caption{{(color online) SARPES data obtained with $hv=24$ eV at $E_{b}=0.15$ eV. (a) Measured MDC in the $\bar{\Gamma} \bar{K}$ direction, showing also the fitted peaks contributing to the intensity. The solid line is the intensity fit. (b) Measured (symbols) and fitted (solid lines) spin polarization data corresponding to (a). (c) The measured $z$ component of the spin polarization vector (symbols) obtained from an azimuthal scan at $k_{||}=0.38$ \AA$^{-1}$, where band $l4$ is located, shows approximately a sine like behavior with $2\pi/3$ periodicity (solid line) and extrema in the $\bar{\Gamma} \bar{K}$ direction. The inset visualizes the out of plane rotation as a function of the azimuthal angle. (d) Spin polarization resulting from a MDC (not shown) in the $\bar{\Gamma} \bar{M}$ direction. The four peaks $l4-l1$ contribute to the spin polarization spectrum. Note the different $k_{||}$ scale between (b) and (d). The insets in (b) and (d) show the corresponding spin polarization components. The symbols refer to those defined in (a). The largest statistical errors of the spin polarization spectra are given by $\Delta$.}}
\label{Fig8}
\end{center}
\end{figure*}

For the Pb/Ag(111) surface alloy, the spin splitting found in the experiment \cite{pacile} is considerably smaller than for the Bi/Ag(111) system. First-principles calculations \cite{bihlmayer} find four bands, derived from two different surface states, with a band structure and band symmetries similar to the Bi/Ag(111) system, and a spin splitting of comparable size. The main difference is that, because Pb has one valence electron less than Bi, the band crossing of the spin split surface states is above the Fermi level in the Pb/Ag(111) surface alloy, which complicates the experimental classification of the different bands. Therefore it has been argued that the inner two bands could originate from the different surface states and are thus not spin split counterparts \cite{bihlmayer}. However, with the knowledge of the band symmetries and the additional information of the spin polarization it is possible to identify and classify the individual bands.
\newline
The measured band structure is displayed in Fig. \nolinebreak \ref{Fig7}, which shows two spin integrated cuts through the SBZ in the $\bar{\Gamma} \bar{K}$ direction for $hv=21.2$ eV and $hv=24$ eV. It illustrates the labeling of the bands and gives their approximate positions. The rightmost band labeled $u$ in Fig. \nolinebreak \ref{Fig7} is again the surface umklapp band, which is only drawn at the right hand side, where it is more intense. The umklapp band is close to the bands $l4$ and $l3$ and can only be separated with the additional spin information.

In Fig. \nolinebreak \ref{Fig8} we give an overview of the results on the Pb/Ag(111) surface alloy. The data are all obtained with the photon energy $hv=24$ eV at the binding energy $E_{b}=0.15$ eV, with the results from the vectorial spin analysis summarized in Table \ref{Table2}. Other photon energies lead to altered intensity distributions due to strong final state effects but do not yield any additional information on the spin structure. From the fit to the momentum distribution curve in the $\bar{\Gamma} \bar{K}$ direction we extract the positions, widths and intensities of the different surface state bands of the Pb/Ag(111) surface alloy. The MDC and the corresponding peaks are shown in Fig. \nolinebreak \ref{Fig8} \nolinebreak (a). The application of the vectorial spin analysis to the spin polarization spectra shown in Fig. \nolinebreak \ref{Fig8} (b) confirms that  $l1/r2$ and $l2/r1$ are spin-split counterparts resulting from the surface state with mostly $sp_{z}$ symmetry, based on their primarily in-plane spin polarization. The splitting is smaller than in the Bi/Ag(111) surface alloy, i.e. 0.05 \nolinebreak \AA$^{-1}$ vs. 0.16 \nolinebreak \AA$^{-1}$, respectively. According to Eq. \ref{alpha} and Eq. \ref{Erashba}, this leads to $\alpha_{R}\nolinebreak=\nolinebreak2.36$ \nolinebreak eV\AA\ and $E_{R}\nolinebreak=\nolinebreak63.9$ \nolinebreak eV using $m^{*}\nolinebreak=\nolinebreak-0.15$ \nolinebreak $m_{e}$ \cite{pacile}. This finding corroborates the assumption that the surface corrugation in the Pb/Ag(111) surface alloy is smaller than for Bi/Ag(111), as suggested in Ref. \cite{bihlmayer}, because this would reduce the spin splitting.
\newline
The assumption of a smaller corrugation is further supported by the spin analysis on the bands $l4$ and $l3$. If the surface corrugation is reduced, these bands will primarily have $m_{j}\nolinebreak=\nolinebreak1/2$ character and similar to Bi/Ag(111) their spin polarization vector will have a significant out-of-plane component in the $\bar{\Gamma} \bar{K}$ direction. This is exactly what is found for the bands $l4$ and $l3$. The out-of-plane rotation is larger for $l4$ than for $l3$ but reduced compared to Bi/Ag(111), i.e. 46$^{\circ}$ and 29$^{\circ}$ for $l4$ and $l3$, respectively.
\newline
Similar to the Bi/Ag(111) surface alloy, adjacent $\Gamma$K directions are not equivalent with respect to $P_{z}$. The dependence of the out-of-plane spin polarization component of band $l4$ at $k_{||}\nolinebreak=\nolinebreak0.38$ \nolinebreak \AA$^{-1}$ on the azimuthal angle is shown in Fig. \nolinebreak \ref{Fig8} (c). It shows an approximate sine like behavior with a $2\pi/3$ periodicity comparable to Fig. \nolinebreak \ref{Fig6} \nolinebreak (b).
\newline
The analysis of the MDC spin polarization spectra from Fig. \nolinebreak \ref{Fig8} (d), to which the bands $l4-l1$ contribute, indicates that the polarization vectors of the bands $l4$ and $l3$ lie in the surface plane for $\bar{\Gamma} \bar{M}$ while the bands are still 100 percent spin polarized: the vanishing spin polarization in the $z$ component is compensated by the appearance of spin polarization in the $y$ component. Combining the information from Fig. \nolinebreak \ref{Fig8} (b), (c), and (d) leads us to the conclusion that the spin polarization vector of band $l4$ rotates out of and into the surface plane as a function of the azimuthal angle, while the in-plane part of the spin polarization vector remains approximately tangential to the constant energy surface. The out-of-plane rotation is schematically indicated by the vectors in Fig. \nolinebreak \ref{Fig8} (c). 
\newline
Again it should be noted that, due to the time inversion symmetry the surface remains non magnetic, which means that the vector sum of all spin polarization vectors throughout the SBZ is zero. For $P_{z}$, this is exemplified by the change of sign for adjacent $\bar{\Gamma} \bar{K}$ directions. Furthermore, from the peak positions of $l2$ and $l1$ (see Table \ref{Table2} for values), we find that the Rashba splitting of the inner two bands is larger for the $\bar{\Gamma} \bar{M}$ direction than for $\bar{\Gamma} \bar{K}$, which is a ramification of the non circular constant energy surface.

\begin{table}[htb]
\begin{center}
\begin{tabular}{c  c  c  c  c  c  c }
\hline
\multicolumn{7}{c}{Pb/Ag(111)} \\
\hline
\multicolumn{7}{c}{$E_{b}=0.15$ eV, $\bar{\Gamma} \bar{K}$} \\
$k_{||}$ (\AA$^{-1}$)& -0.38 & -0.35 & -0.23 & -0.18 & 0.18 & 0.23 \\
$\phi$ $(^\circ)$& -42 $\pm32$ & 120 $\pm21$ & -85 $\pm2$ & 93 $\pm2$ & -81 $\pm2$ & 102 $\pm3$ \\
$\theta$ $(^\circ)$& -47 $\pm8$ & 28 $\pm11$ & -1 $\pm2$ & 3 $\pm2$ & 0 $\pm2$ & 1 $\pm3$ \\
\hline
 & $l4$ & $l3$ & $l2$ & $l1$ & $r1$ & $r2$ \\
 \hline
\multicolumn{7}{c}{$E_{b}=0.15$ eV, $\bar{\Gamma} \bar{M}$} \\
$k_{||}$ (\AA$^{-1}$)& -0.40 & -0.36 & -0.27 & -0.18 &  &  \\
$\phi$ $(^\circ)$& -83 $\pm10$ & 124 $\pm7$ & -83 $\pm2$ & 100 $\pm2$ &  & \\
$\theta$ $(^\circ)$& 4 $\pm11$ & 2 $\pm8$ & -1 $\pm2$ & -2 $\pm2$ &  &  \\
\hline
\end{tabular}
\end{center}
\caption{Band positions and directions (given by $\theta$ and $\phi$) of the spin polarization vectors of the surface state bands $l4-r2$ for the Pb/Ag(111) surface alloy at $E_{b}=0.15$ eV. The angular errors result from the fitting procedure to the spin polarization spectra.}
\label{Table2}
\end{table}

\section{Conclusions}
In this work, a novel method to analyze spin resolved photoemission data was introduced and applied to the Bi/Ag(111) and Pb/Ag(111) surface alloys. We have shown that this approach can yield information that is not accessible by spin integrated photoemission. It can determine the three-dimensional spin polarization vectors of individual bands and resolve otherwise undistinguishable bands through the utilization of the spin as an additional tag. Moreover, the method is robust against strong intensity variations due to matrix element effects, because it references the spin polarization contribution of each individual band to the measured peak intensity.
\newline
We have confirmed recent experimental and theoretical interpretations regarding the giant Rashba splitting in these systems and observed a large out-of-plane spin polarization component.
Furthermore, we have shown that the Rashba type spin-split bands are completely spin polarized, and that changes in the measured spin polarization for different crystallographic directions are due to variations in the directions of the spin polarization vectors, band positions and their intensity distribution and not due to changes in the length of the spin polarization vector.

\begin{acknowledgements}
Technical support by C. Hess, F. Dubi and M. Kl\"ockner is gratefully acknowledged. The measurements have been performed at the Swiss Light Source, Paul Scherrer Institut, Villigen, Switzerland. 
\end{acknowledgements}



\begin{thebibliography}{}
\bibitem{sens}
The different detection efficiencies of the opposite channels $I_{\hat{\alpha}}^{i}$ are accounted for with empirically determined sensitivity factors.
\end{thebibliography}
\end{document}